%% file: main.tex
\documentclass[12 pt]{article}
\usepackage[margin=1in]{geometry}
\usepackage[affil-it]{authblk}
\usepackage[table]{xcolor}
\usepackage{hyperref}
\hypersetup{
    colorlinks=true,
    linkcolor=magenta,
    filecolor=magenta,      
    urlcolor=blue,
    citecolor = blue,
}
\usepackage{amsmath,amssymb,amsfonts}
\usepackage{mathtools}
\usepackage{enumerate}
\usepackage{algorithmic}
\usepackage{graphicx}
\usepackage{textcomp}
\usepackage{xcolor}
\usepackage{sistyle}
\usepackage[
backend=biber,
style=chicago-authordate,
citestyle=chicago-authordate
]{biblatex}
\input{preamble.tex}
\SIthousandsep{,}
\def\BibTeX{{\rm B\kern-.05em{\sc i\kern-.025em b}\kern-.08em
    T\kern-.1667em\lower.7ex\hbox{E}\kern-.125emX}}
\begin{document}
\title{Social Contagion and Associative Diffusion in Multilayer Network\\
}

\author{Heng-Chien Liou\thanks{b05901023@ntu.edu.tw; liouhengchien@gmail.com\\
}, Hsuan-Wei Lee
}
\affil{Department of Electrical Engineering,
National Taiwan University\\
Institute of Sociology, Academia Sinica 
}

\maketitle

\begin{abstract}
    The question that how cultural variation emerges has drawn lots of interest in sociological inquiry. Sociologists predominantly study such variation through the lens of social contagion, which mostly attributes cultural variation to the underlying structural segregation, making it epiphenomenal to the pre-existing segregated structure. On the other hand, arguing culture doesn't spread like a virus, an alternative called associative diffusion was proposed, in which cultural transmission occurs not at the preference of practices, but at the association between practices. The associative diffusion model then successfully explains cultural variation without attributing it to a segregated social structure. The contagion model and associative diffusion model require different types of relationships and interactions to make cultural transmission possible. In reality, both types of relationships exist most of the time. In light of this concern, we proposed combining the two models with the multilayer network framework. On one layer, agents casually observed the behaviors of others, updating their belief about the association between practices, just like the associative diffusion model; on another layer, agents' preference of practices are directly influenced by closed others, just like the social contagion model. In the meantime, the constraint satisfaction between preference and association is used to link the update of both, thereby making each individual a coherent entity in terms of preference and association. Using this approach, we entangle the effect of social contagion and associative diffusion through multilayer networks. For the baseline, we explore the model dynamics on three common network models: fully connected, small-world, and scale-free. The results show nontrivial dynamics between the two extremes of the contagion model and the associative diffusion model, justifying our claim that it is necessary to consider the two models at the same time. 
\end{abstract}

\section{Introduction}

How culture emerges might be one of the most fundamental problems in social science. Most of the grand sociological theorists directly or indirectly address this problem. For \textcite{durkheim1995elementary}, the clues are hidden in totems and rituals; for \textcite{weber1978economy}, the interpretations of individuals play important roles in such processes. Along with these theoretical attempts, empirical and simulational approaches also contribute to our understanding of basic cultural processes. 

Since the invention of computers, these machines have been important tools for humans; sociologists are no exception. Sociologists have adopted the use of computers for sociological inquiries for decades, under the name of mathematical sociology and computational sociology \parencite{coleman1964introduction, sorensen1978mathematical, macy2002factors}. With the help of increasingly advanced computational power, sociologists can test various hypotheses and theories about society \textit{in silico}. Along with other important questions in sociology, the inquiry about cultural emergence greatly benefits from the computational approach. 

In the questioning of cultural emergence and transmission, the concept of \textit{social contagion} is possibly one of the most popular. The epidemiological metaphor is easily understood and used, and it has lead to various interesting discoveries. However, under the pure assumption of contagious transmission, underlying structural segregation is needed for cultural variation. Such a requirement renders cultural variation epiphenomenal to pre-existing social structure. 

Addressing this issue, \textcite{goldberg2018beyond} formulate another model called \textit{associative diffusion}. Such a model assumes the perception of associations between behaviors, instead of the preference of behaviors, is transmitted in the social interaction. Within this model, even with a fully-mixed, homogeneous population, cultural variation can emerge. 

In this study, we ask and try to answer a rather simple question, ``what if both processes exist and function in a society''. To answer this question, we introduce the concept of multilayer networks \parencite{bianconi2018multilayer}, or multiplex networks, to our theorizing framework. Such a framework allows us to specify the channel for each separate process, thereby studying the phenomena under different conditions. Similar to the precedent, we use computer simulation as our main method to approach this question.

\section{How Does Culture Transmit, Emerge, and Differentiate?}

\subsection{Social Contagion}

The paradigm of social contagion is possibly the most dominant approach in the theorizing of cultural emergence, particularly for those using computation as their main approach for social science inquires. 

In the social contagion paradigm, beliefs or behaviors are ``transmitted'' in the networked population. People adopt certain opinions, behaviors, and beliefs because others do so. This framework rest on human's innate tendency to mimic others' behaviors and beliefs. The reasons behind involve various factors, such as the need to resolve uncertainty and the evolutionary drive to stay in the group. 
These kinds of dyadic, microscopic influences are expected to be able to drive the macroscopic cultural changes in society. Although this trend of social contagion analysis is relatively recent, the history of this kind of contagious theorizing can be traced back to the work of \textcite{de1903laws}, one of the earliest sociological theorist. 

With the advance of computational methods and the inputs from physics and applied mathematics to social science, various kinds of contagion models have been proposed. Scholars from various disciplines together have already studied these models intensively and characterized models' behaviors under different kinds of conditions. Social scientists also have examined the empirical foundations of various models and validate their applicability through the data collected from the real world. For example, \textcite{christakis2013social} conducted extensive studies on the potential spread of various kinds of traits, behaviors, health conditions, and unified them under the framework of ``Social Contagion Theory''.

Although the virus-like transmission has become the mainstream of the contagion model, sociologists also work on other variants. For example, \textcite{centola2007complex} introduced a more refined idea of the complex contagion. In the complex contagion model, in contrast with the simple contagion, people adopt behaviors or opinions only if a certain proportion or number of their neighbors have already adopted. Using this model, \citeauthor{centola2007complex} found dramatically different system dynamics from those in the simple contagion model. 

Yet, even with diverse variants, contagion models still rely on pre-existing structural segregation to explain cultural differentiation \parencite{goldberg2018beyond}. Thus, an alternative model to account for cultural variation in the homogeneous population has been proposed.

\subsection{Associative Diffusion}

Arguing culture doesn't spread like a virus, \textcite{goldberg2018beyond} proposed an alternative called \textit{associative diffusion}. They question the central assumption of the contagion model that cultural transmission is possible only through viable and meaningful relationships. Instead, they focus on the conductivity of superficial interaction. The proposed model draws inspiration from cognitive science, distinguishing between adapting a cultural practice and interpreting it. Instead of the direct transmission of practice adoption in the previous virus-like models, the perception of associations between behaviors is diffused through the population.   

In the associative diffusion model, cultural transmission follows a two-stage process. Observing others enacting certain behaviors, agents update their perception about what behaviors are compatible with one another, and then further adjust their preference by random perturbation, to cohere with the association. 

\begin{table}[!t]
    \centering
    \caption{The overview of associative diffusion model, adopted from  \cite{goldberg2018beyond}. }
    \begin{tabular}{l}
        \hline 
        {Initialization}  \\[0.5ex] 
        \hline
        Each agent holds two types of information: \\
        1. association: $\mb{R}_{ij} = 1, \forall i, j \in \mcal{K}$, \\
        2. preference: $\bs{V}_i \sim \mrm{Unif} \lp  -1, 1\rp$. \\
        [0.5ex] 
        \hline 
        {Association Transmission} \\[0.5ex] 
        \hline
        Select agent $B \in \mcal{V}$ at random, and then select another agent $A \in \mcal{N}_{\text{out}} \lp B \rp$. \\
        $B$ observes $A$ exhibiting practices $i$ and $j$ with probabilities $\msf{P} \lp i \rp$ and $\msf{P} \lp j \rp$. \\
        $B$ updates $\mb{R}_{ij} = \mb{R}_{ij} +1$. \\
        $B$ updates preferences, $\bs{V}'$ , with $\bs{V}'_k = \bs{V}_k +  \sim \mrm{N} \lp 0, 1 \rp  $, where $k=\mrm{argmin} \lbp \lv \bs{V}_i \rv, \lv \bs{V}_j \rv \rbp$. \\
        If $\mrm{CS} \lp \bs{V}', \mb{R} \rp > \mrm{CS} \lp \bs{V}, \mb{R} \rp$, $\bs{V}'$ is retained, otherwise revert to $\bs{V}$.
    \end{tabular}
    \label{tab:golldber2018}
\end{table}

More formally, each agent can be associated with a node in the network, and the links between describe the possible channels for interaction. Each agent holds two types of information, association $\mb{R}$ and preference $\bs{V}$: association $\mb{R} \in \mbb{R}^{K \times K}$ is a matrix describing the perceived pairwise association among $K$ kinds of practices; preference $\bs{V} \in \mbb{R}^{K}$ is a vector representing the agent's preference over the $K$ practices of interest. 

During the model evolution, in each round, one of the agent $B$ is randomly selected from the network, and then another agent $A $ is selected from $\mcal{N}_{\text{out}} \lp B \rp$, the set of out-neighbors of $B$. $A$ exhibits practices $i$ and $j$ with probability $\msf{P} \lp i \rp,  \msf{P} \lp j \rp  \propto \exp \lp V_i \rp, \exp \lp V_j \rp$. Observing $A$'s exhibition, B update the association matrix $\mb{R}$ with $\mb{R}_{ij} = \mb{R}_{ij} + 1$. And then $B$ further update the weaker of the two preference with $ \Delta v \sim \mrm{N} \lp 0, 1 \rp$. $B$ retains the update of preference if and only if the it is more consistent with the association matrix $\mb{R}$. \citeauthor{goldberg2018beyond} proposed to calculate this consistency with a function called ``constraint satisfaction'', $\mrm{CS}$, with the formula 

\begin{align}
    \mrm{CS} \lp \bs{V}, \mb{R} \rp = \frac{2}{K \lp K-1\rp} \sum_{i=1}^K \sum_{j=1}^K \lv \mb{R}_{ij} - \mb{\Omega}_{ij} \rv, \quad \text{where } \mb{\Omega}_{ij} = \lv \bs{V}_i - \bs{V}_j \rv
\end{align}

An overview of the associative diffusion model is presented in Table \ref{tab:golldber2018}. \citeauthor{goldberg2018beyond} further studied the model behavior under various kinds of network topology: 2-agents network, fully-connected 30-agents network, scale-free networks, and small-world networks. Cultural variation naturally emerges from these various kinds of networks, without attributing it to a segregated social structure.

\subsection{A Combination of the Two Paradigms}

In the contagion model, it is assumed that cultural transmission relies on enduring and meaningful relationships to provide the necessary bandwidth and incentive to adopt certain behaviors or opinions; in the associative diffusion model, those superficial interactions are expected to drive cultural variation. In reality, both kinds of relationships and interactions exist. And we can expect, if the assumptions behind both models are justifiable, these two kinds of processes most likely will co-occur. And their effects will intertwine. 

In this study, we synthesize the modeling approaches of these two seemingly competing paradigms. We assume that in a given population, there are two kinds of channels to transmit information. Through one kind of channel, agents can transmit the information of association; through another kind, agents transmit their preference. Under our assumptions, these two kinds of transmission processes drive the cultural evolution in society. 

To grasp the basic idea, we can imagine the first kind of channels as constituted by superficial interactions with others, and the second kind of channels as maintained by more meaningful relationships. Living in the same district might be sufficient to constitute the first kind of channel, and sharing close friendships can provide both kinds of channels at the same time.

In this formulation, the first kind of transmission is directly equivalent to the transmission in the associative diffusion model, and the second kind of transmission can be viewed as a sub-type of the contagion process. Therefore, in this hypothetical setting, we can study how the interplay of these two models determine cultural evolution. 
\section{Modeling Framework and Measurement}

\subsection{The Multilayer Network Framework}

With regards to the need to describe populations with two different means of cultural transmission, we use the conceptual framework of the multilayer network, or its subclass multiplex network, to formulate our model. The notations used are mostly consistent with \gentextcite{bianconi2018multilayer}, with slight modifications in font styles. 

We consider a population in which, the pattern of interaction can be describes as a multiplex network $\mcal{M}$, 

\begin{align}
    \mcal{M} = \lp \mcal{Y}, \vec{G} \rp, \quad \text{where } \mcal{Y} = \lbp 1, 2 \rbp, \vec{G} = \lp G_1, G_2 \rp.
\end{align}

$\mcal{Y}$ is the index set of layers, and $\vec{{G}}$ is the ordered list of networks describing the interaction within each layer. $G_{a} = \lp  \mcal{V}_{a}, \mcal{E}_{a} \rp$, where $\mcal{V}_a$ is the set of nodes on layer $a$, and $\mcal{E}_a$ is the set of edges on layer $a$, respectively. In our case, each network $G_{a} = \lp  \mcal{V}_{a}, \mcal{E}_{a} \rp$ is formed by same set of nodes, 

\begin{align}
    \mcal{V}_1 = \mcal{V}_2 = \mcal{V} =  \lbp 1, 2, \ldots, N \rbp.
\end{align}
This special case of multiplex networks with $\lv \mcal{Y} \rv = 2$ is also called a duplex network \parencite{bianconi2018multilayer}.

In our framework, each node $A$ in set $\mcal{V}$ represents an agent in the population. $\mcal{E}_1$ and $\mcal{E}_2$ respectively describe two sets of interaction existing in this population. It is important to note that, at this stage, we have no restriction on the relationship between $\mcal{E}_1$ and $\mcal{E}_2$; for example, $\mcal{E}_1 \cap \mcal{E}_2 = \varnothing$ is not necessarily true.
Unless otherwise specified, the networks we considered are directed \parencite{newman2018networks}; that is, the relationship between arbitrary two can be asymmetric. The fact that $A$ can transmit information to $B$ doesn't guarantee that $B$ can transmit information to $A$, and vice versa.

\subsection{Model}

\begin{table}[!t]
    \centering
    \caption{Model overview}
    \begin{tabular}{l}
        \hline 
        {Initialization}  \\[0.5ex] 
        \hline
        Each agent holds two types of information: \\
        1. association: $\mb{R}_{ij} = 1, \forall i, j \in \mcal{K}$, \\
        2. preference: $\bs{V}_i \sim \mrm{Unif} \lp  -1, 1\rp$. \\
        [0.5ex] 
        \hline 
        {Association Transmission}, with probability $1-\alpha$  \\[0.5ex] 
        \hline
        Select agent $B \in \mcal{V}$ at random, and then select another agent $A \in \mcal{N}^{\lb 1 \rb}_{\text{out}} \lp B \rp$. \\
        $B$ observes $A$ exhibiting practices $i$ and $j$ with probabilities $\msf{P} \lp i \rp$ and $\msf{P} \lp j \rp$. \\
        $B$ updates $\mb{R}_{ij} = \mb{R}_{ij} +1$. \\
        $B$ updates preferences, $\bs{V}'$ , with $\bs{V}'_k = \bs{V}_k +  \sim \mrm{N} \lp 0, 1 \rp  $, where $k=\mrm{argmin} \lbp \lv \bs{V}_i \rv, \lv \bs{V}_j \rv \rbp$. \\
        If $\mrm{CS} \lp \bs{V}', \mb{R} \rp > \mrm{CS} \lp \bs{V}, \mb{R} \rp$, $\bs{V}'$ is retained, otherwise revert to $\bs{V}$.\\
        [0.5ex] 
        \hline 
        {Preference Transmission}, with probability $\alpha$  \\[0.5ex] 
        \hline
        Select agent $B \in \mcal{V}$ at random, and then select another agent $A \in \mcal{N}^{\lb 2 \rb}_{\text{out}} \lp B \rp$. \\
        $B$ observes $A$ exhibiting practices $i$ with probabilities $P \lp i \rp$. \\
        $B$ updates preference $\bs{V}_i = \bs{V}_i + 1$.\\
        $B$ updates associations, $\mb{R}'$ , with $\mb{R}'_{i, *} = \mb{R}_{i, *} +  \sim \mrm{N} \lp \bs{0}, \mb{I} \rp  $. \\
        If $\mrm{CS} \lp \bs{V}, \mb{R}' \rp > \mrm{CS} \lp \bs{V}, \mb{R} \rp$, $\mb{R}'$ is retained, otherwise revert to $\mb{R}$.
    \end{tabular}
    \label{tab:my_model}
\end{table}

In our model, there are two necessary components. One is the modeling sequence of transmitting association, another is the modeling of transmitting preference, corresponding to the process of associative diffusion model and contagion model, respectively. 

To make the experimental results comparable to \gentextcite{goldberg2018beyond}, we make the process of transmitting association identical to the modeling sequence of \citeauthor{goldberg2018beyond}'s associative diffusion model, which is shown in the ``Association Transmission'' part of Table \ref{tab:my_model}. On the other hand, we design the process of transmitting preference, so that the process is analogous to the process of transmitting association, which is shown in the ``Preference Transmission'' part of Table \ref{tab:my_model}.

To be more specific, being the same as the associative diffusion model, each agent in $\mcal{V}$ is associated with two kinds of information: association matrix $\mb{R}$ and preference vector $\bs{V}$. In each round, a random agent $B$ will be select to update the information associated. 

With probability $1-\alpha$, agent $B$ will observe another agent $A$ in the out-neighborhood  $\mcal{N}_{\text{out}}^{[1]} \lp B\rp$, defined by layer $1$, or edge set $\mcal{E}_1$. Agent $A$ will practice two behaviors $i, j$, according to a probability function determined by preference vector. Agent $B$ will then update the association matrix $\mb{R}$ with $\mb{R}_{ij} = \mb{R}_{ij}+1$, which can be viewed as agent $A$ transmitting the information of association to agent $B$. And then agent B randomly modifies preference vector $\bs{V}$ to increase constraint satisfaction. The procedure is almost the same as the associative diffusion model, except for specifying the neighborhood by layer $1$. 

On the contrary, with probability $\alpha$, agent $B$ will observe another agent $A$ in the out-neighborhood  $\mcal{N}_{\text{out}}^{[2]} \lp B\rp$, defined by layer $2$, or edge set $\mcal{E}_2$. Agent $A$ will practice only one behavior $i$. Agent $B$ will then update the preference vector $\bs{V}$ with $\bs{V}_{i} = \bs{V}_{i}+1$, which can be viewed as agent $A$ transmit the information of preference to agent $B$. And then agent B randomly modify association matrix $\mb{R}$ to increase constraint satisfaction. The procedure is a complement of the associative diffusion model, where we interchange the roles of $\bs{V}$ and $\mb{R}$ with corresponding modifications. 

A complete overview of the proposed model is depicted in Table \ref{tab:my_model}.

\subsection{Measurement}

For comparison purposes, most measurements we used are directly adopted from those in \gentextcite{goldberg2018beyond}. 

\paragraph{Evaluative Agreement}

The first measure we can consider is the agreement among agents' preference, about how they evaluate different cultural practices. Following \citeauthor{goldberg2018beyond}'s argument, there are two kinds of measure, \textit{Preference Similarity} and \textit{Preference Congruence}. For preference similarity, we simply measure how similar it is between agents' evaluations, that is, $\rho \lp \bs{V}, \bs{V}^* \rp$, the Pearson correlation between two agents' preference vectors \parencite{wasserman2013all}. For preference congruence, we measure whether agents' preferences follow the same pattern, which is mathematically defined as $\lv \rho \lp \bs{V}, \bs{V}^* \rp \rv$, the absolute value of the Pearson correlation. We then average across all possible pairs of distinct agents to result in group-level measures.

\paragraph{Interpretive Agreement} \citeauthor{goldberg2018beyond} measure interpretive agreement between two agents as the distance between their normalized association matrices, $\mb{R}$ and $\mb{R}*$, with the formula as  $\lv \lv \mb{R}, \mb{R}^* \rv \rv = \lp 1/{K^2} \rp \sum_{i=1}^K \sum_{j=1}^K \lv \wtild{\mb{R}}_{ij} - \wtild{\mb{R}}^*_{ij} \rv$, where $\wtild{\mb{R}} = \mb{R}/\mrm{max} \lp \mb{R} \rp$. In this study, we instead  operationalize interpretive agreement as $\rho \lp \mb{R}, \mb{R}^* \rp$, the element-wise Pearson correlation between matrices $ \mb{R} $ and $\mb{R}^*$, following the calculation about evaluative agreement previously mentioned. We coin this measure as \textit{association similarity}, in parallel with preference similarity.

\paragraph{Behavioral Predictability} In contrast with previous two paragraphs where we discuss the agreement between different agents, as an inter-agent, dyad-level measure, we now consider the measure within agent itself. We want to know that how observing agent's one behavior can help us predict the occurrence of another behavior. \citeauthor{goldberg2018beyond} used the measure of mutual information to characterize this predictability, as $\mrm{I} = \sum_{i, j} \msf{P} \lp i, j \rp \log \frac{\msf{P} \lp i, j \rp}{\msf{P} \lp i \rp \msf{P} \lp j\rp}$, a common measure used in information theory \parencite{cover2006elements}.

\paragraph{Cultural differentiation} \citeauthor{goldberg2018beyond} operationalize cultural differentiation as the optimal number of preference clusters. If all agents' preference vectors belong to the same cluster, there is no cultural differentiation; on the other hand, for any number larger than one, cultural differentiation exists. The optimal number of clusters is estimated by a procedure called gap statistics \parencite{tibshirani2001estimating} and its variants \parencite{mohajer2011comparison}, with adaptation to make the procedure suitable for the preference vector considered \parencite{goldberg2018beyond}. Due to the limit in computational resources, we have not been able to report the results of this measure yet. 

\section{Experiments and Results}

\begin{figure}[!t]
\begin{subfigure}[b]{0.425\textwidth}
\includegraphics[trim=30 0 25 20, clip, width=1\linewidth]{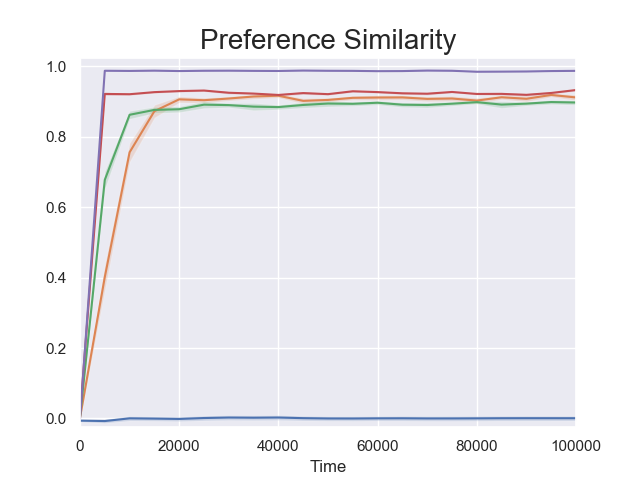} 
\caption{Preference Similarity}
\label{fig:fullcopy_preference}
\end{subfigure}
\begin{subfigure}[b]{0.425\textwidth}
\includegraphics[trim=30 0 25 20, clip, width=1\linewidth]{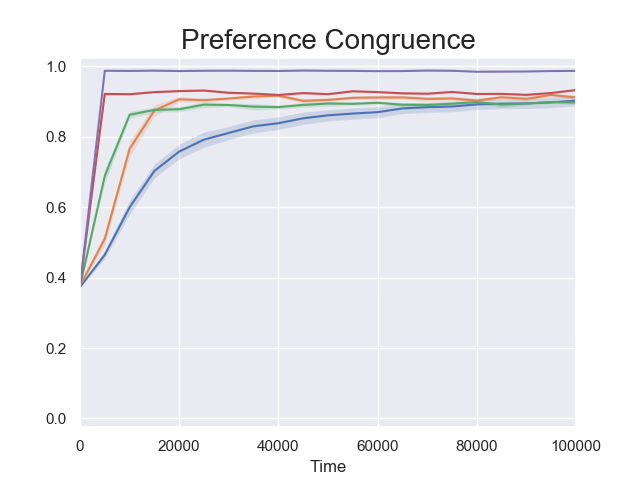}
\caption{Preference Congruence}
\label{fig:fullcopy_evaluative}
\end{subfigure}
\quad
\begin{subfigure}[b]{0.10\textwidth}
\includegraphics[width=\linewidth]{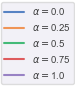}
\end{subfigure}
\vskip\baselineskip
\begin{subfigure}[b]{0.425\textwidth}
\includegraphics[trim=30 0 25 20, clip, width=1\linewidth]{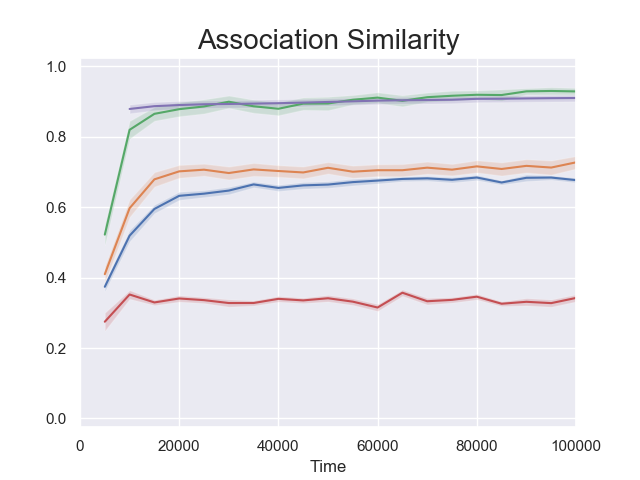} 
\caption{Association Similarity}
\label{fig:fullcopy_interpretive}
\end{subfigure}
\begin{subfigure}[b]{0.425\textwidth}
\includegraphics[trim=30 0 25 20, clip, width=1\linewidth]{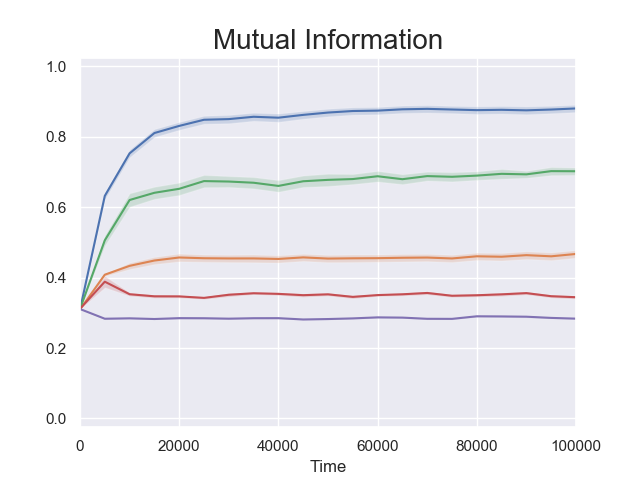}
\caption{Mutual Information}
\label{fig:fullcopy_meaningful}
\end{subfigure}

\caption{Simulation results of model on network with duplicated fully-connected layers}
\label{fig:fullcopy}
\end{figure}

\begin{figure}[!t]
\begin{subfigure}[b]{0.425\textwidth}
\includegraphics[trim=30 0 25 20, clip, width=1\linewidth]{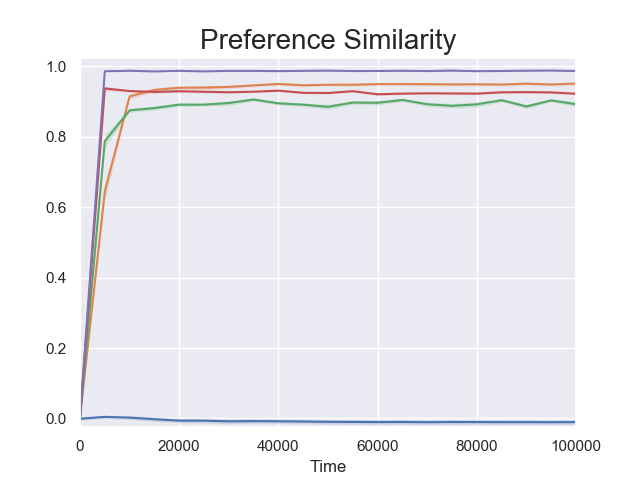} 
\caption{Preference Similarity}
\label{fig:scalefreecopy_preference}
\end{subfigure}
\begin{subfigure}[b]{0.425\textwidth}
\includegraphics[trim=30 0 25 20, clip, width=1\linewidth]{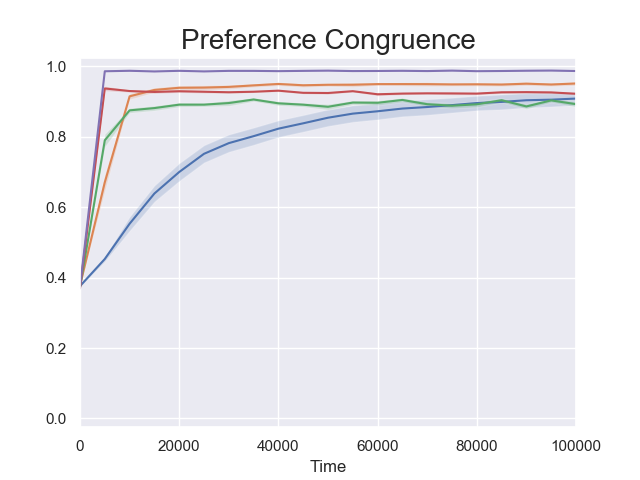}
\caption{Preference Congruence}
\label{fig:scalefreecopy_evaluative}
\end{subfigure}
\quad
\begin{subfigure}[b]{0.10\textwidth}
\includegraphics[width=\linewidth]{image/legend.png}
\end{subfigure}
\vskip\baselineskip
\begin{subfigure}[b]{0.425\textwidth}
\includegraphics[trim=30 0 25 20, clip, width=1\linewidth]{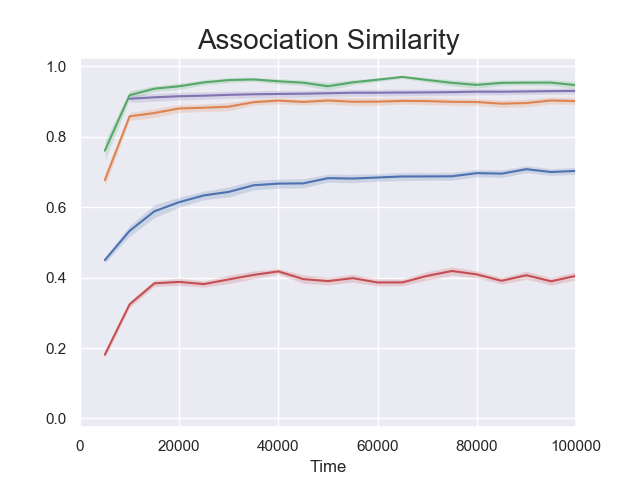} 
\caption{Association Similarity}
\label{fig:scalefreecopy_interpretive}
\end{subfigure}
\begin{subfigure}[b]{0.425\textwidth}
\includegraphics[trim=30 0 25 20, clip, width=1\linewidth]{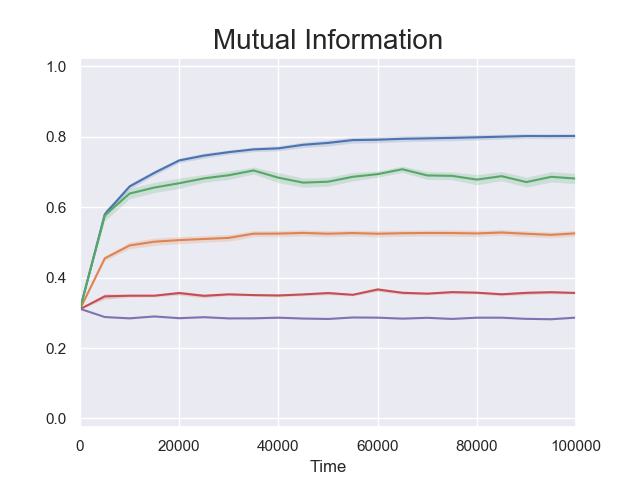}
\caption{Mutual Information}
\label{fig:scalefreecopy_meaningful}
\end{subfigure}

\caption{Simulation results of model on network with duplicated scale-free layers}
\label{fig:scalefreecopy}
\end{figure}

\begin{figure}[!t]
\begin{subfigure}[b]{0.425\textwidth}
\includegraphics[trim=30 0 25 20, clip, width=1\linewidth]{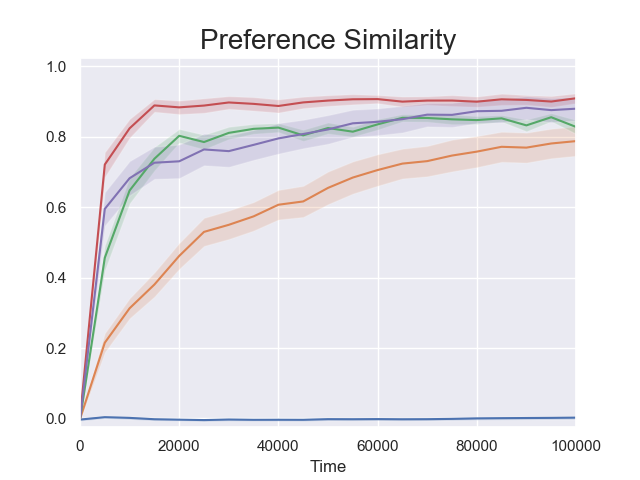} 
\caption{Preference Similarity}
\label{fig:smallworldcopy_preference}
\end{subfigure}
\begin{subfigure}[b]{0.425\textwidth}
\includegraphics[trim=30 0 25 20, clip, width=1\linewidth]{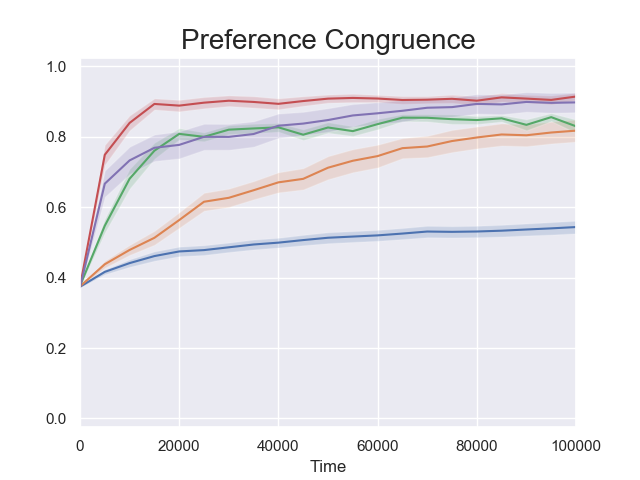}
\caption{Preference Congruence}
\label{fig:smallworldcopy_evaluative}
\end{subfigure}
\quad
\begin{subfigure}[b]{0.10\textwidth}
\includegraphics[width=\linewidth]{image/legend.png}
\end{subfigure}
\vskip\baselineskip
\begin{subfigure}[b]{0.425\textwidth}
\includegraphics[trim=30 0 25 20, clip, width=1\linewidth]{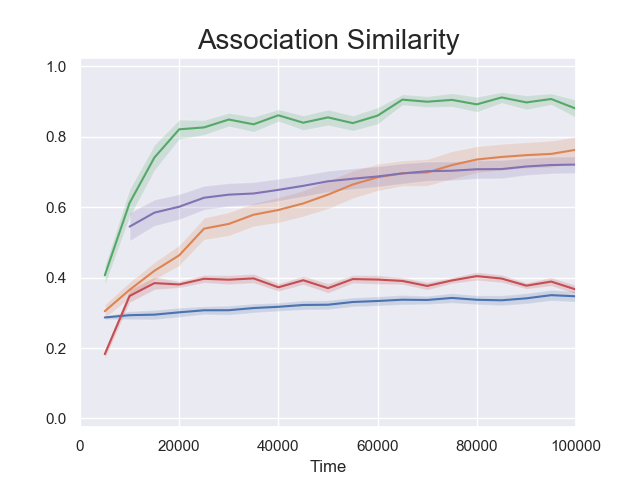} 
\caption{Association Similarity}
\label{fig:smallworldcopy_interpretive}
\end{subfigure}
\begin{subfigure}[b]{0.425\textwidth}
\includegraphics[trim=30 0 25 20, clip, width=1\linewidth]{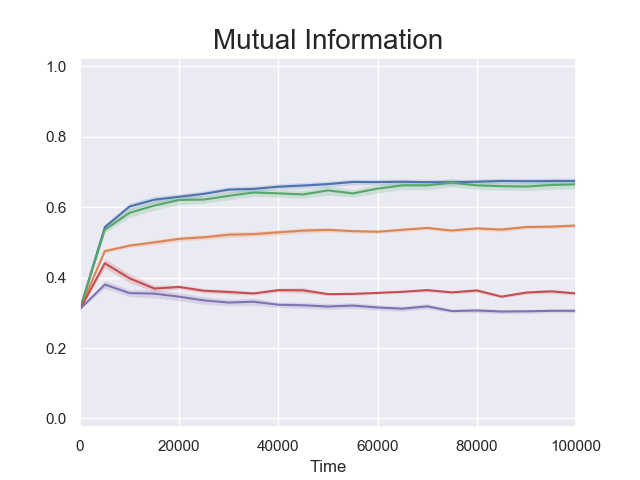}
\caption{Mutual Information}
\label{fig:smallworldcopy_meaningful}
\end{subfigure}

\caption{Simulation results of model on network with duplicated small-world layers}
\label{fig:smallworldcopy}
\end{figure}

\begin{figure}[!t]
\begin{subfigure}[b]{0.4\textwidth}
\includegraphics[trim=30 0 25 20, clip, width=1\linewidth]{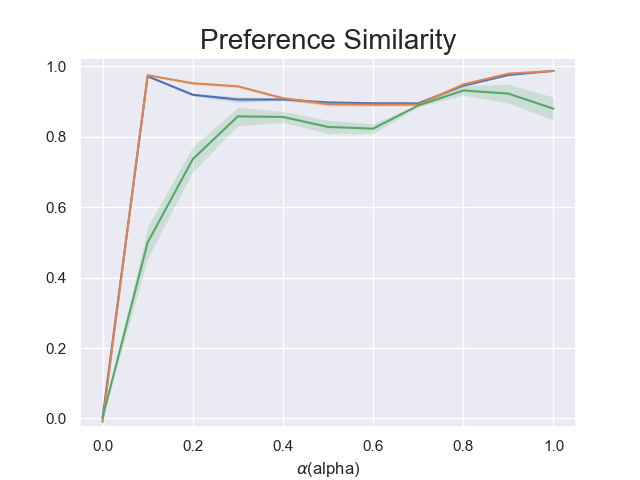} 
\caption{Preference Similarity}
\label{fig:alphacombined_preference}
\end{subfigure}
\begin{subfigure}[b]{0.4\textwidth}
\includegraphics[trim=30 0 25 20, clip, width=1\linewidth]{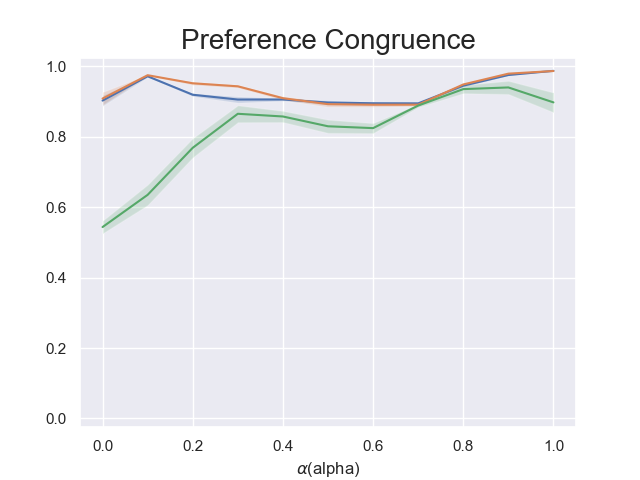}
\caption{Preference Congruence}
\label{fig:alphacombined_evaluative}
\end{subfigure}
\quad
\begin{subfigure}[b]{0.15\textwidth}
\includegraphics[width=\linewidth]{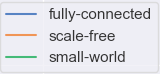}
\end{subfigure}
\vskip\baselineskip
\begin{subfigure}[b]{0.4\textwidth}
\includegraphics[trim=30 0 25 20, clip, width=1\linewidth]{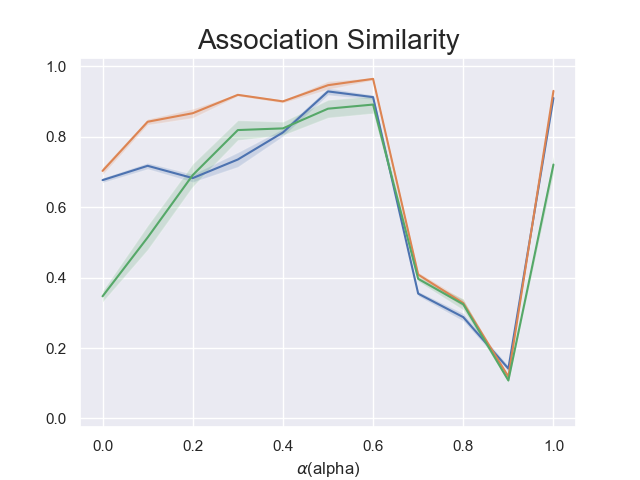} 
\caption{Association Similarity}
\label{fig:alphacombined_interpretive}
\end{subfigure}
\begin{subfigure}[b]{0.4\textwidth}
\includegraphics[trim=30 0 25 20, clip, width=1\linewidth]{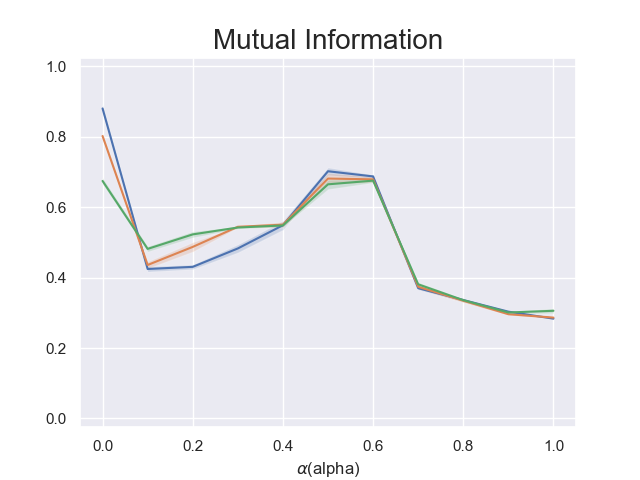}
\caption{Mutual Information}
\label{fig:alphacombined_meaningful}
\end{subfigure}

\caption{The final values across different $\alpha$ for the three networks considered.}
\label{fig:alphacombined}
\end{figure}

In this section, we explore the characteristics of our model. We start from the most simple case of networks with duplicate layers and then progress into more complicated cases. 

Unless otherwise specified, all the experiments are conducted through computer simulation on a laptop. The simulation is implemented in Python3 programming language \parencite{10.5555/1593511}, with module Graph-tool for standard graph manipulation \parencite{peixoto_graph-tool_2014}. Due to the limit in computational resources, we take necessary steps to lighten the load, for example, lowering the number of simulations, sampling data for visualization, without drastically invalidating our results.

In this short paper, the first scenario we considered is for the cases where $\mcal{E}_1 = \mcal{E}_2 = \mcal{E}$. Under such a condition, the model considered can be reduced to a model on equivalent single-layer networks. 

\subsection{Networks with Duplicated Fully-connected Layers}

Following the exploration of \citeauthor{goldberg2018beyond}, we start with fully-connected networks. In fully-connected networks, or complete graph, each node is connected to every other node. And for directed network, each node points to every other node, or using the set notation, $\mcal{N}_{\text{out}} \lp A \rp = \mcal{V} - A, \forall A \in \mcal{V}$. A fully-connected network is maximally homogeneous, with each node occupying the same role in the network structure. \citeauthor{goldberg2018beyond} have shown that given a fully-connected graph, associative diffusion model can drive the emergence of cultural variation. In this part of study, we simulate our models, a combination of associative diffusion model and contagion model, on fully-connected graph with 30 agents ($\lv \mcal{V} \rv = 30, \lv \mcal{E} \rv = 870$) to see how the dynamics differ from what \citeauthor{goldberg2018beyond} got. The results for different values of $\alpha$'s are shown in Figure \ref{fig:fullcopy}, and we can discuss them in turns. 

When $\alpha = 0.0$, in which the model is reduced to \citeauthor{goldberg2018beyond}'s associative diffusion model, agents are divided into two clusters. Within each cluster, they have a similar preference, and between clusters, the preference follows a similar pattern but in opposite ways. This result can be seen in the near-zero value of preference similarity in Figure \ref{fig:fullcopy_preference}, and the continuously growing preference congruence in Figure \ref{fig:fullcopy_evaluative}. 

For $\alpha > 0$, a qualitative change happens for preference similarity: it is far away from zero. This is true whenever contagion has come to play. And there is no longer easily observable differences in preference similarity and congruence, which implies that all agent follows a similar pattern in preference, with slight variation. And when $\alpha = 1.0$, both preference similarity and congruence are approximately one, which shows the complete prevalence results from the contagion effect. 

Both preference similarity and congruence remain relatively stable within the range about $\lb 0.9, 1 \rb$ after $t= 20,000$, for $\alpha > 0$. And the two measures no longer provide useful distinctions between different values of $\alpha$. In contrast, the value of association similarity and mutual information after stabilizing remain distinguishable. And the relationship between $\alpha$ and stabilizing value of mutual information and interpretive similarity is far from trivial, which indicates that although the modeling sequence is a linear combination of the two models, the resulting nonlinear dynamics deserve a brand-new analysis.

Next, we consider the other two kinds of topology used by \citeauthor{goldberg2018beyond}: the scale-free network and the small-world network. Both types of networks are commonly used in network science research \parencite{newman2018networks}.

\subsection{Networks with Duplicated Scale-free Layers}

A scale-free network, or power-law network, is a network with in-degree following a power law distribution, $P_{\text{in}} \lp k \rp \propto k^{-\gamma}$. This property is common among technological networks, such as power grid and Internet \parencite{newman2018networks}. For social networks, these kinds of networks are often used to model networks constituted by superficial interactions \parencite{goldberg2018beyond}.  The simulation result on graph with 30 agents ($\gamma = 3, k_{\text{out}} = 6, \lv \mcal{V} \rv = 30, \lv \mcal{E} \rv = 180$) are shown in Figure \ref{fig:scalefreecopy}. 

A close inspection into Figure \ref{fig:scalefreecopy} reveals that most of our discussions about fully-connected networks can directly apply. The relative positions of different curves are astonishingly similar to the relative positions of those of fully-connected networks. Minor numerical differences are in expectation: since the network topology is drastically different (for example, $\lv \mcal{E} \rv = 180$ in scale-free networks, and $\lv \mcal{E} \rv = 870$ in fully-connected networks), there's no way we would expect them to behave exactly the same. 

The only exception occurs for interpretive similarity at $\alpha = 0.25$. In fully-connected layers, interpretive similarity curve at $\alpha = 0.25$ is right next to the curve at $\alpha = 0.0$, while in scale-free layers, the curve is more close with the curve at $\alpha = 1.0$. We haven't found a reasonable explanation yet.

\subsection{Networks with Duplicated Small-world Layers}

A small-world network is a network with high clustering and short distances between arbitrary nodes, two commonly held assumptions in the study of social networks. Compared to scale-free networks, small-world networks are often used to describe those networks with segregated nature, such as the relationship among families, tribes, or friends \parencite{goldberg2018beyond}.

The results with 30 agents and 5 clusters ($k_{\text{out}} = 5, k_{\text{in}} = 5, \lv \mcal{V} \rv = 30, \lv \mcal{E} \rv = 150$) shown in Figure \ref{fig:smallworldcopy} reveal several interesting points about dynamics on small-world networks. As indicated by the discussion in \citeauthor{goldberg2018beyond}'s work, the pre-segregated structure can prevent the graph to be overwhelmed by contagion dynamics, as shown by the curve of preference similarity and congruence for $\alpha = 1.0$. Interestingly, different from what is shown in fully-connected networks and scale-free networks, in which the preference similarity at $\alpha = 1.0$ rises as the fastest to the maximum, in small-world networks, the curve of $\alpha = 1.0$ climbs relatively slower, and is overtaken by both the curve of $\alpha = 0.5$ and $\alpha = 0.75$. This observation suggest that introducing associative diffusion into contagion process can help speed up the spread of dominant norms in segregated social networks.

\subsection{A Final Comparison}

At the end of the experiment section, we present a comparison across the three types of networks we consider. All graphs consist of 30 agents ($\lv \mcal{V} \rv = 30$), but of different number of edges and degree distribution. The result is shown in Figure \ref{fig:alphacombined}, in which we get the final value (at $t = 100,000$) of each measure, plotted with various value of $\alpha$. Through these plots, we can quickly grasp how the selection of $\alpha$ influence modeling results, and how the model behaves on different types of networks. 

As observed in Figure \ref{fig:alphacombined_preference} and \ref{fig:alphacombined_evaluative}, the preference similarity and congruence for small-world networks are both lower than the other two, while fully-connected and scale-free networks behave almost the same for these two measures. This is consistent with the general understanding that small-world networks can impede the spread. On the other hand, the almost identical results for fully-connected and scale-free networks are astonishing, since they differ largely in several ways; for example, the number of edges is 870 and 180, respectively. 

In contrast, the above statement cannot apply to the other two measures in Figure \ref{fig:alphacombined_interpretive} and \ref{fig:alphacombined_meaningful}, in which there's no clear order of magnitude. The non-monotonic correspondence of $\alpha$ and modeling results is also easily observable from the plots.

\section{Discussion anc Conclusion}

In this paper, we only include the results on networks with duplicated layers, with three kinds of basic topology: fully-connected, scale-free, and small-world. In fact, the proposed model can be used on all networks with two layers, not necessarily duplicated, and there are lots of networks remained to explore.

For our next step, we will complete the current study by investigating the cultural variation in these various settings, which is the same as estimating the optimal number of clusters for preference vectors. These results will provide an intuitive indicator of how the culture evolves in these scenarios.

We will then explore the model dynamics when two layers are generated by the same but uncorrelated random process, for example, two uncorrelated scale-free layers. And we can also study the dynamics on networks with layers generated by different random processes, for example, with $G_1$ generated as scale-free layers, while $G_2$ is generated as small-world layers.

We can even further study the model dynamics on networks with correlated but nonidentical random layers. For example, the networks generated by two-layer preferential attachment \parencite{bianconi2018multilayer}. With the models and analytical tools from the multilayer network community, we can characterize how the proposed model will behave in these various circumstances. 

In this work, we present a model to combine the popular contagion paradigm and the newly proposed associative diffusion paradigm. The model results in complicated nontrivial dynamics, which is far from the interpolation of the two separate models. We also show that the combination of evaluative and interpretive influence speed up the spread in networks with segregated structure, which is shown in the results of small-world networks. This paper also demonstrates the unique advantages of using computer simulation for social inquiry. We hope that this work can stimulate more sociological practitioners to embrace the use of computation in their inquiries about the social world.

\section*{Appendix}

For clarity and comprehensibility in the formulation and mathematical description, we summarize the notations used throughout the paper in Table \ref{tab:notations}.

\begin{table}[!ht]
    \centering
    \caption{Notaions}
    \begin{tabular}{|l|l|l|}
    \multicolumn{1}{c}{Category} & \multicolumn{1}{c}{Font and Style} & \multicolumn{1}{c}{Example} \\
    \hline
    Scalar/Index /Agent    & normal italic serif  & $i, j; A, B$ \\ 
    \hline 
    Vector & boldface italic serif & $\bs{V}$ \\
    \hline 
    Matrix & boldface upright serif & $\mb{R}$ \\
    \hline
    Set/Function returning set   & caligraphic & $\mcal{V}, \mcal{E}; \mcal{N}_\text{out} \lp  B\rp$ \\
    \hline
    Probability mass function & sans serif & $\msf{P} \lp i \rp$ \\
    \hline
    Distribution & normal upright serif & $\mrm{Unif}, \mrm{N}$ \\
    \hline
    \end{tabular}
    \label{tab:notations}
\end{table}

\printbibliography

\end{document}

%% file: preamble.tex
\usepackage{amsmath,amssymb,mathrsfs}
\usepackage{amsthm}
\usepackage{epsfig,epsf,graphicx,graphics}
\usepackage{url}
\usepackage{array}
\usepackage{multicol}
\usepackage{empheq}
\usepackage{subcaption}

\newcommand{\lp}{\left(}
\newcommand{\rp}{\right)}
\newcommand{\lb}{\left[}
\newcommand{\rb}{\right]}

\newcommand{\lv}{\left\vert}
\newcommand{\rv}{\right\vert}
\newcommand{\lbp}{\left\{}
\newcommand{\rbp}{\right\}}

\newcommand{\mcal}{\mathcal}

\newcommand{\mrm}{\mathrm}

\newcommand{\wtild}{\widetilde}
\newcommand{\mb}{\mathbf}
\newcommand{\mbb}{\mathbb}
\newcommand{\bs}{\boldsymbol}
\newcommand{\msf}{\mathsf}

\renewcommand\thesection{\Alph{section}}
\renewcommand\thesubsection{\arabic{subsection}}
\renewcommand\thesubsubsection{\alph{subsubsection})}

\allowdisplaybreaks

%% file: ref.bib
@article{goldberg2018beyond,
  title={Beyond social contagion: Associative diffusion and the emergence of cultural variation},
  author={Goldberg, Amir and Stein, Sarah K},
  journal={American Sociological Review},
  volume={83},
  number={5},
  pages={897--932},
  year={2018},
  publisher={SAGE Publications Sage CA: Los Angeles, CA}
}

@book{de1903laws,
  title={The laws of imitation},
  author={Tarde, Gabriel},
  year={1903},
  publisher={H. Holt}
}

@book{durkheim1995elementary,
  title = {The elementary forms of religious life},
  author = {Durkheim, Emile},
  translator = { Karen Fields},
  date = {1995},
  origdate = {1912},
  publisher = {The Free Press}
  }

@book{weber1978economy,
  title={Economy and society: An outline of interpretive sociology},
  author={Weber, Max},
  volume={1},
  year={1978},
  publisher={Univ of California Press}
}

@book{coleman1964introduction,
  title={Introduction to mathematical sociology.},
  author={Coleman, James Samuel},
  year={1964}
}

@article{sorensen1978mathematical,
  title={Mathematical models in sociology},
  author={S{\o}rensen, Aage B},
  journal={Annual review of sociology},
  volume={4},
  number={1},
  pages={345--371},
  year={1978},
  publisher={Annual Reviews 4139 El Camino Way, PO Box 10139, Palo Alto, CA 94303-0139, USA}
}

@article{macy2002factors,
  title={From factors to actors: Computational sociology and agent-based modeling},
  author={Macy, Michael W and Willer, Robert},
  journal={Annual review of sociology},
  volume={28},
  number={1},
  pages={143--166},
  year={2002},
  publisher={Annual Reviews 4139 El Camino Way, PO Box 10139, Palo Alto, CA 94303-0139, USA}
}

@book{bianconi2018multilayer,
  title={Multilayer networks: structure and function},
  author={Bianconi, Ginestra},
  year={2018},
  publisher={Oxford university press}
}

@article{christakis2013social,
  title={Social contagion theory: examining dynamic social networks and human behavior},
  author={Christakis, Nicholas A and Fowler, James H},
  journal={Statistics in medicine},
  volume={32},
  number={4},
  pages={556--577},
  year={2013},
  publisher={Wiley Online Library}
}

@book{newman2018networks,
  title={Networks},
  author={Newman, Mark},
  year={2018},
  publisher={Oxford university press}
}

@article{peixoto_graph-tool_2014,
         title = {The graph-tool python library},
         url = {http://figshare.com/articles/graph_tool/1164194},
         doi = {10.6084/m9.figshare.1164194},
        %  urldate = {2014-09-10},
         journal = {figshare},
         author = {Peixoto, Tiago P.},
         year = {2014},
         keywords = {all, complex networks, graph, network, other}}

@book{10.5555/1593511, 
 author = {Van Rossum, Guido and Drake, Fred L.}, 
 title = {Python 3 Reference Manual}, 
 year = {2009}, 
 isbn = {1441412697}, 
 publisher = {CreateSpace}, 
 address = {Scotts Valley, CA} 
}

@article{tibshirani2001estimating,
  title={Estimating the number of clusters in a data set via the gap statistic},
  author={Tibshirani, Robert and Walther, Guenther and Hastie, Trevor},
  journal={Journal of the Royal Statistical Society: Series B (Statistical Methodology)},
  volume={63},
  number={2},
  pages={411--423},
  year={2001},
  publisher={Wiley Online Library}
}

@article{centola2007complex,
  title={Complex contagions and the weakness of long ties},
  author={Centola, Damon and Macy, Michael},
  journal={American journal of Sociology},
  volume={113},
  number={3},
  pages={702--734},
  year={2007},
  publisher={The University of Chicago Press}
}

@book{wasserman2013all,
  title={All of statistics: a concise course in statistical inference},
  author={Wasserman, Larry},
  year={2013},
  publisher={Springer Science \& Business Media}
}

@book{cover2006elements,
  title={Elements of information theory},
  author={Cover, Thomas M. and Thomas, Jay A.},
  edition = {2},
  year={2006},
  publisher={John Wiley \& Sons}
}

@article{mohajer2011comparison,
  title={A comparison of Gap statistic definitions with and without logarithm function},
  author={Mohajer, Mojgan and Englmeier, Karl-Hans and Schmid, Volker J},
  journal={arXiv preprint arXiv:1103.4767},
  year={2011}
}
